\documentclass[twocolumn,pra,showpacs,preprintnumbers]{revtex4}
\usepackage{graphicx}
\begin{document}
\title{Stationary Josephson effect in a weak-link between nonunitary
triplet superconductors}
\author{G. Rashedi$^{1,3}$ and Yu. A.
Kolesnichenko$^{2}$}
\address{$^1$ Institute for Advanced Studies in Basic Sciences,
 Zanjan, 45195-1159, Iran\\ $^2$ B.Verkin Institute for Low
Temperature Physics
 Engineering of National Academy of Sciences of Ukraine, 47,
  Lenin ave , 61103, Kharkov, Ukraine\\
$^3$ Department of Physics, Faculty of Science, University of
Shahrekord, Shahrekord, P.O.Box 115, Iran}

\date{\today}

\begin{abstract}
A stationary Josephson effect in a weak-link between misorientated
nonunitary triplet superconductors is investigated theoretically.
The non-self-consistent quasiclassical Eilenberger equation for
this system has been solved analytically. As an application of
this analytical calculation, the current-phase diagrams are
plotted for the junction between two nonunitary bipolar $f-$wave
superconducting banks. A spontaneous current parallel to the
interface between superconductors has been observed. Also, the
effect of misorientation between crystals on the Josephson and
spontaneous currents is studied. Such experimental investigations
of the current-phase diagrams can be used to test the pairing
symmetry in the above-mentioned superconductors.
\end{abstract}
\maketitle
\section{Introduction}
In recent years, the triplet superconductivity has become one of
the modern subjects for researchers in the field of
superconductivity \cite {Ueda,Maeno,Mackenzie}. Particularly, the
nonunitary spin triplet state in which Cooper pairs may carry a
finite averaged intrinsic spin momentum has attracted much
attention in the last decade \cite{Tou,Machida}. A triplet state
in the momentum space $\mathbf{k}$ can be described by the order
parameter
${\hat{\Delta}}(\mathbf{k})=i(\mathbf{d}(\mathbf{k})\cdot {\hat{
\sigma}})\hat{\sigma}_{y}$ in a 2$\times $2 matrix form in which
$\hat{\sigma}_{j}$ are 2$\times $2 Pauli matrices $\left(
{\hat{\sigma}=}\left(
\hat{\sigma}_{x},\hat{\sigma}_{y},\hat{\sigma}_{z}\right) \right)
$. The three dimensional complex vector $\mathbf{d}(\mathbf{k)}$
(gap vector) describes the triplet pairing state. In the
nonunitary state, the product
${\hat{\Delta}}(\mathbf{k}){\hat{\Delta}}(\mathbf{k})^{\dagger
}=\mathbf{d}(\mathbf{k})\cdot \mathbf{d}^{\ast
}(\mathbf{k})+i(\mathbf{d}(\mathbf{k})\times \mathbf{d}^{\ast
}(\mathbf{k}))\cdot {\hat{\sigma}}$ is not a multiple of the unit
matrix. Thus in a non-unitary state the time reversal symmetry is
necessarily broken spontaneously and a spontaneous moment
$\mathbf{m}(\mathbf{k})=i\mathbf{d}(\mathbf{k})\times
\mathbf{d}^{\ast }(\mathbf{k})$ appears at each point $\mathbf{k}$
of the momentum space. In this case the macroscopically averaged
moment $<\mathbf{m}(\mathbf{k})>$ integrated on the Fermi surface
does not vanish. The value $\mathbf{m}(\mathbf{k})$ is related to
the net spin average by $Tr[{\hat{\Delta}}(\mathbf{k})^{\dag
}{\hat{\sigma}_{j}}{\hat{\Delta}}(\mathbf{k})]$. It is clear that
the total spin average over the Fermi surface can be nonzero. As
an application, the nonunitary bipolar state of $f-$wave pairing
symmetry has been considered for the $B-$phase of
superconductivity in the $UPt_{3}$ compound which has been created
at low temperatures $T$ and small values of the magnetic field $H$
\cite{Machida,Ohmi}.

In the present paper, the ballistic Josephson weak link via an
interface between two superconducting bulks with different
orientations of the crystallographic axes is investigated. This
type of weak link structure can be used for the demonstration of
the pairing symmetry in the superconducting phase
\cite{Stefanakis}. Consequently, we generalize the formalism of
paper \cite{Mahmoodi} for the weak link between triplet
superconducting bulks with a nonunitary order parameter. In the
paper \cite {Mahmoodi}, the Josephson effect in the point contact
between unitary $f-$wave triplet superconductors has been studied.
Also, the effect of misorientation on the charge transport has
been investigated and a spontaneous current tangential to the
interface between the $f-$wave superconductors, has been observed.

In this paper the nonunitary bipolar $f-$wave model of the order
parameter is considered. It is shown that the current-phase
diagrams are totally different from the current-phase diagrams of
the junction between the unitary triplet ( axial and planar)
$f-$wave superconductors \cite{Mahmoodi} . Roughly speaking, these
different characters can be used to distinguish between nonunitary
bipolar $f-$wave superconductivity and the other types of
superconductivity. In the weak-link structure between the
nonunitary $f-$wave superconductors, the spontaneous current
parallel to the interface has been observed as a fingerprint for
unconventional superconductivity and spontaneous time reversal
symmetry breaking. The effect of misorientation on the spontaneous
and Josephson currents is investigated. It is possible to find the
value of the phase difference in which the Josephson current is
zero but the spontaneous current, which is produced by the
interface and is tangential to the interface, is present. In some
configurations and at the zero phase difference, the Josephson
current is not generally zero but has a finite value. This finite
value corresponds to a spontaneous phase difference which is
related to the misorientation between the gap vectors$\mathbf{d}$.

The arrangement of the rest of this paper is as follows. In
Sec.\ref{section2} we describe the configuration that we have
investigated. For a non-self-consistent model of the order
parameter, the quasiclassical Eilenberger equations
\cite{Eilenberger} are solved and suitable Green functions have
been obtained analytically. In Sec.\ref{section3} the formulas
obtained for the Green functions have been used for the
calculation of the current densities at the interface. An analysis
of numerical results will be presented in Sec.\ref{section4}
together with some conclusions in Sec.\ref{section5}.
\section{Formalism and Basic Equations}
\label{section2} We consider a model of a flat interface $y=0$
between two misorientated nonunitary $f-$wave superconducting
half-spaces (Fig.1) as a ballistic Josephson junction. In the
quasiclassical ballistic approach, in order to calculate the
current, we use ``transport-like'' quasiclassical Eilenberger
equations \cite{Eilenberger} for the energy integrated Green
functions $\breve{g}\left(
\mathbf{\hat{v}}_{F},\mathbf{r},\varepsilon _{m}\right) $
\begin{equation}
\mathbf{v}_{F}\cdot \nabla \breve{g}+\left[ \varepsilon
_{m}\breve{\sigma} _{3}+i\breve{\Delta},\breve{g}\right] =0,
\label{Eilenberger}
\end{equation}
and the normalization condition $\breve{g}\breve{g}=\breve{1}$,
where $\varepsilon _{m}=\pi T(2m+1)$ are discrete Matsubara
energies $m=0,1,2,...,$ $T$ is the temperature, $\mathbf{v}_{F}$
is the Fermi velocity and
$\breve{\sigma}_{3}=\hat{\sigma}_{3}\otimes \hat{I}$ in which
$\hat{\sigma} _{j}\left( j=1,2,3\right) $ are Pauli matrices.
\begin{figure}[tbp]
\includegraphics[width=0.9\columnwidth]{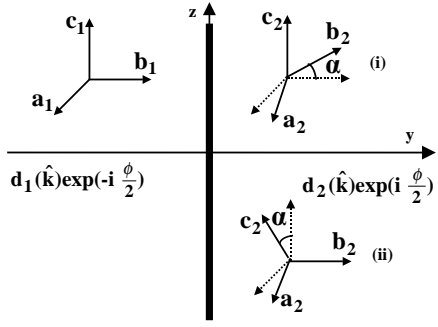}
\caption{Scheme of a flat interface between two superconducting
bulks which are misorientated as much as $\protect\alpha $.}
\label{fig1}
\end{figure}
The Matsubara propagator $\breve{g}$ can be written in the form:
\begin{equation}
\breve{g}=\left(
\begin{array}{cc}
g_{1}+\mathbf{g}_{1}\cdot \mathbf{\hat{\sigma}} & \left(
g_{2}+\mathbf{g} _{2}\cdot \hat{\sigma}\right) i\hat{\sigma}_{2} \\
i\hat{\sigma}_{2}\left( g_{3}+\mathbf{g}_{3}\cdot
\hat{\sigma}\right)  & g_{4}-\hat{\sigma}_{2}\mathbf{g}_{4}\cdot
\hat{\sigma}\hat{\sigma}_{2}
\end{array}
\right),\label{Green's function}
\end{equation}
where the matrix structure of the off-diagonal self energy
$\breve{\Delta}$ in the Nambu space is
\begin{equation}
\breve{\Delta}=\left(
\begin{array}{cc}
0 & \mathbf{d}\cdot \hat{\sigma}i\hat{\sigma}_{2} \\
i\hat{\sigma}_{2}\mathbf{{d^{\ast }}\cdot \hat{\sigma}} & 0
\end{array}
\right).\label{order parameter}
\end{equation}
The nonunitary states, for which $\mathbf{d\times d} ^{\ast }\neq
0,$ are investigated. Fundamentally, the gap vector (order
parameter) $\mathbf{d}$ has to be determined numerically from the
self-consistency equation \cite{Ueda}, while in the present paper,
we use a non-self-consistent model for the gap vector which is
much more suitable for analytical calculations \cite{Kulik}.
Solutions to Eq. (\ref{Eilenberger}) must satisfy the conditions
for the Green functions and the gap vector $\mathbf{d}$ in the
bulks of the superconductors far from the interface as follow:
\begin{equation}
\breve{g}=\frac{1}{\Omega _{n}}\left(
\begin{array}{cc}
\varepsilon _{m}(1-\mathbf{A}_{n}\cdot \mathbf{\hat{\sigma}}) &
\left[i\mathbf{d}_{n}-\mathbf{d}_{n}\times \mathbf{A}_{n}\right]
\cdot
\hat{\sigma}i\hat{\sigma}_{2} \\
i\hat{\sigma}_{2}\left[ i\mathbf{d}_{n}^{\ast
}+\mathbf{d}_{n}^{\ast }\times \mathbf{A}_{n}\right] \cdot
\hat{\sigma} & -\varepsilon \hat{\sigma}_{2}(1+\mathbf{A}_{n}\cdot
\hat{\sigma})\hat{\sigma}_{2}
\end{array}
\right)   \label{Bulk solution}
\end{equation}
where
\begin{equation}
\hspace{-0.4cm}\mathbf{A}_{n}=\frac{i\mathbf{d}_{n}\times
\mathbf{d}_{n}^{\ast }}{\varepsilon _{m}^{2}+\mathbf{d}_{n}\cdot
\mathbf{d}_{n}^{\ast }+\sqrt{(\varepsilon
_{m}^{2}+\mathbf{d}_{n}\cdot \mathbf{d}_{n}^{\ast
})^{2}+(\mathbf{d}_{n}\times \mathbf{d}_{n}^{\ast })^{2}}}
\end{equation}
and
\begin{equation}
\hspace{-0.55cm}\Omega _{n}=\sqrt{\frac{2[(\varepsilon
_{m}^{2}+\mathbf{d} _{n}\cdot \mathbf{d}_{n}^{\ast
})^{2}+(\mathbf{d}_{n}\times \mathbf{d} _{n}^{\ast
})^{2}]}{\varepsilon _{m}^{2}+\mathbf{d}_{n}\cdot \mathbf{d}
_{n}^{\ast }+\sqrt{(\varepsilon _{m}^{2}+\mathbf{d}_{n}\cdot
\mathbf{d} _{n}^{\ast })^{2}+(\mathbf{d}_{n}\times
\mathbf{d}_{n}^{\ast })^{2}}}}
\end{equation}
\begin{equation}
\mathbf{d}\left( \pm \infty \right) =\mathbf{d}_{2,1}\left(
T,\mathbf{\hat{v}}_{F}\right) \exp \left( \mp \frac{i\phi
}{2}\right) \label{Bulk order parameter}
\end{equation}
where $\phi $ is the external phase difference between the order
parameters of the bulks and $n=1,2$ label the left and right half
spaces respectively. It is clear that poles of the Green function
in the energy space are in
\begin{equation} \Omega _{n}=0.
\end{equation}
Consequently,
\begin{equation}
(-E^{2}+\mathbf{d}_{n}\cdot \mathbf{d}_{n}^{\ast
})^{2}+(\mathbf{d}_{n}\times \mathbf{d}_{n}^{\ast })^{2}=0
\end{equation}
and
\begin{equation}
E=\pm \sqrt{\mathbf{d}_{n}\cdot \mathbf{d}_{n}^{\ast }\pm
i\mathbf{d} _{n}\mathbf{\times d}_{n}^{\ast }}
\end{equation}
in which $E$ is the energy value of the poles. The Eq.
(\ref{Eilenberger}) has to be supplemented by the continuity
conditions at the interface between superconductors. For all
quasiparticle trajectories, the Green functions satisfy the
boundary conditions both in the right and left bulks as well as at
the interface. The system of equations (\ref{Eilenberger}) and the
self-consistency equation for the gap vector $\mathbf{d}$
\cite{Ueda} can be solved only numerically.
\begin{figure}[tbp]
\includegraphics[width=0.9\columnwidth]{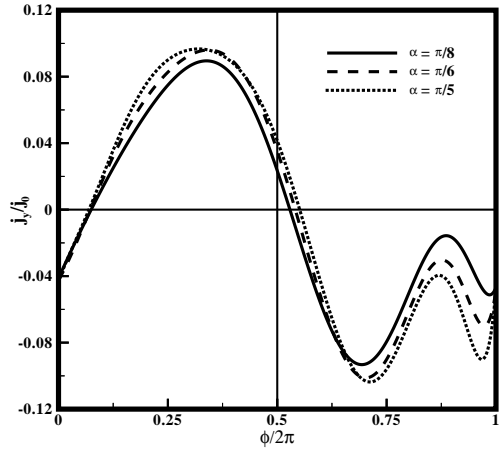}
\caption{Component of the current normal to the interface
(Josephson current) versus the phase difference $\protect\phi $
for the junction between nonunitary bipolar $f-$wave bulks ,
$T/T_{c}=0.15$, geometry (i) and different misorientations.
Currents are given in units of $j_{0}=\frac{\protect\pi
}{2}eN(0)v_{F}\Delta _{0}(0).$} \label{fig2}
\end{figure}
For unconventional superconductors such solution requires the
information about the interaction between the electrons in the
Cooper pairs and the nature of unconventional superconductivity in
novel compounds which in most cases are unknown. Also, it has been
shown that the absolute value of a self-consistent order parameter
is suppressed near the interface and at the distances of the order
of the coherence length, while its dependence on the direction in
the momentum space almost remains unaltered \cite{Barash}. This
suppression of the order parameter changes the amplitude value of
the current, but does not influence the current-phase dependence
drastically. For example, it has been verified in
Refs.\cite{Coury} for the junction between unconventional $d$-wave
superconductors, in Ref.\cite{Barash} for the case of unitary
``$f$-wave'' superconductors and in Ref.\cite{Viljas} for pinholes
in $^{3}He,$ that there is good qualitative agreement between
self-consistent and non-self-consistent results for not very large
angles of misorientation. It has also been observed that the
results of the non-self-consistent model in \cite{Yip} are similar
to experiment \cite{Backhaus}. Consequently, despite the fact that
this solution cannot be applied directly to a quantitative
analysis of a real experiment, only a qualitative comparison of
calculated and experimental current-phase relations is possible.
In our calculations, a simple model of the constant order
parameter up to the interface is considered and the pair breaking
and scattering on the interface are ignored. We believe that under
these strong assumptions our results describe the real situation
qualitatively. In the framework of such a model, the analytical
expressions for the current can be obtained for a certain form of
the order parameter.
\section{Analytical results}\label{section3}
The solution of Eq. (\ref{Eilenberger}) allows us to calculate the
current densities. The expression for the current is:
\begin{equation}
\mathbf{j}\left( \mathbf{r}\right) =2i\pi eTN\left( 0\right)
\sum_{m}\left\langle \mathbf{v}_{F}g_{1}\left(
\mathbf{\hat{v}}_{F},\mathbf{r},\varepsilon _{m}\right)
\right\rangle
 \label{charge-current}
\end{equation}
where $\left\langle ...\right\rangle $ stands for averaging over
the directions of an electron momentum on the Fermi surface
${\mathbf{\hat{v}}} _{F},$ and $N\left( 0\right) $ is the electron
density of states at the Fermi level of energy.
\begin{figure}[tbp]
\includegraphics[width=0.9\columnwidth]{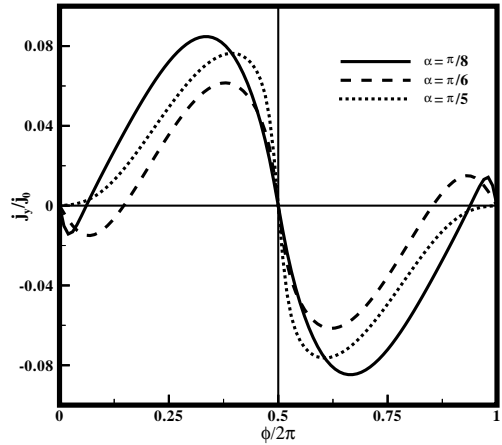}
\caption{Component of the current normal to the interface
(Josephson current) versus the phase difference $\protect\phi $
for the junction between nonunitary bipolar $f-$wave bulks ,
$T/T_{c}=0.15$, geometry (ii) and different misorientations.}
\label{fig3}
\end{figure}
We assume that the order parameter is constant in space and in
each half-space it equals its value (\ref{Bulk order parameter})
far from the interface in the left or right bulks. For such a
model, the current-phase dependence of a Josephson junction can be
calculated analytically. It enables us to analyze the main
features of current-phase dependence for any model of the
nonunitary order parameter. The Eilenberger equations (\ref
{Eilenberger}) for Green functions $\breve{g}$, which are
supplemented by the condition of continuity of solutions across
the interface, $y=0$, and the boundary conditions at the bulks,
are solved for a non-self-consistent model of the order parameter
analytically. In the ballistic case the system of equations for
functions $g_{i}$ and $\mathbf{g}_{i}$ can be decomposed into
independent blocks of equations. The set of equations which
enables us to find the Green function $g_{1}$ is:
\begin{eqnarray}
v_{F}\hat{\mathbf{k}}\nabla g_{1} &=&i\left( \mathbf{d}\cdot
\mathbf{g}_{3}-\mathbf{d}^{\ast }\cdot \mathbf{g}_{2}\right) ;
\label{a} \\
v_{F}\hat{\mathbf{k}}\nabla \mathbf{g}_{-} &=&-2\left(
\mathbf{d\times
g} _{3}+\mathbf{d}^{\ast }\mathbf{\times g}_{2}\right) ;  \label{b} \\
v_{F}\hat{\mathbf{k}}\nabla \mathbf{g}_{2} &=&-2\varepsilon
_{m}\mathbf{g}_{2}+2ig_{1}\mathbf{d}+\mathbf{d}\times
\mathbf{g}_{-}; \label{d} \\ v_{F}\hat{\mathbf{k}}\nabla
\mathbf{g}_{3} &=&2\varepsilon _{m}\mathbf{g}
_{3}-2ig_{1}\mathbf{d}^{\ast }+\mathbf{d}^{\ast }\times
\mathbf{g}_{-}; \label{c}
\end{eqnarray}
where $\mathbf{g}_{-}=\mathbf{g}_{1}-\mathbf{g}_{4}.$ The Eqs.
(\ref{a})-(\ref{d}) can be solved by integrating over the
ballistic trajectories of electrons in the\ right and left
half-spaces. The general solution satisfying the boundary
conditions (\ref{Bulk solution}) at infinity is \begin{equation}
g_{1}^{\left( n\right) }=\frac{\varepsilon _{m}}{\Omega
_{n}}+a_{n}e^{-2s\Omega _{n}t}; \label{e}
\end{equation}
\begin{equation}
\mathbf{g}_{-}^{\left( n\right) }=-2\frac{\varepsilon _{m}}{\Omega
_{n}}\mathbf{A}_{n}+\mathbf{C}_{n}e^{-2s\Omega _{n}t};  \label{f}
\end{equation}
\begin{equation}
\hspace{-0.3cm}\mathbf{g}_{2}^{\left( n\right)
}=\frac{i\mathbf{d}_{n}-\mathbf{d}_{n}\times
\mathbf{A}_{n}}{\Omega _{n}}-\frac{2ia_{n}\mathbf{d}
_{n}+\mathbf{d}_{n}\times \mathbf{C}_{n}}{2s\eta \Omega
_{n}-2\varepsilon _{m}}e^{-2s\Omega _{n}t};  \label{g}
\end{equation}
\begin{equation}
\hspace{-0.3cm}\mathbf{g}_{3}^{\left( n\right)
}=\frac{i\mathbf{d}_{n}^{\ast }+\mathbf{d}_{n}^{\ast }\times
\mathbf{A}_{n}}{\Omega _{n}}+\frac{2ia_{n}\mathbf{d}_{n}^{\ast
}-\mathbf{d}_{n}^{\ast }\times \mathbf{C}_{n}}{2s\eta \Omega
_{n}+2\varepsilon _{m}}e^{-2s\Omega _{n}t};  \label{h}
\end{equation}
where $t$ is the time of flight along the trajectory, $sgn\left(
t\right) =sgn\left( y\right) =s$ and $\eta =sgn\left( v_{y}\right)
.$ By matching the solutions (\ref{e}-\ref{h}) at the interface
$\left( y=0,t=0\right) $, we find constants $a_{n}$ and
$\mathbf{C}_{n}.$ Indices $n=1,2$ label the left and right
half-spaces respectively. The function $g_{1}\left( 0\right)
=g_{1}^{\left( 1\right) }\left( -0\right) =g_{1}^{\left( 2\right)
}\left( +0\right)$ which is a diagonal term of the Green matrix
and determines the current density at the interface, $y=0$, is as
follows:
\begin{equation}
g_{1}\left( 0\right) =\frac{\eta (\mathbf{d}_{2}\cdot
\mathbf{d}_{2}(\eta \Omega _{1}+\varepsilon
)^{2}-\mathbf{d}_{1}\cdot \mathbf{d}_{1}(\eta \Omega
_{2}-\varepsilon )^{2}+B)}{[\mathbf{d}_{2}(\eta \Omega
_{1}+\varepsilon )+\mathbf{d}_{1}(\eta \Omega _{2}-\varepsilon
)]^{2}} \label{charge-term}
\end{equation}
where $B=i\mathbf{d}_{1}\times \mathbf{d}_{2}\cdot
(\mathbf{A}_{1}\mathbf{+A}_{2})(\eta \Omega _{2}-\varepsilon
)(\eta \Omega _{1}+\varepsilon ).$ We consider a rotation
$\breve{R}$ only in the right superconductor (see Fig.1), i.e.,
$\mathbf{d}_{2}(\hat{\mathbf{k}})=\breve{R}\mathbf{d}_{1}(
\breve{R}^{-1}\hat{\mathbf{k}});$ $\hat{\mathbf{k}}$ is the unit
vector in the momentum space.
\begin{figure}[tbp]
\includegraphics[width=0.9\columnwidth]{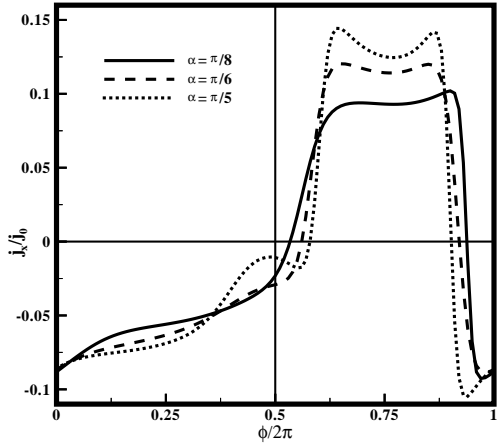}
\caption{The $x-$component of the current tangential to the
interface versus the phase difference $\protect\phi $ for the
junction between nonunitary bipolar $f-$wave superconducting
bulks, $T/T_{c}=0.15$, geometry (i) and the different
misorientations.} \label{fig4}
\end{figure}
The crystallographic $c$-axis in the left half-space is selected
parallel to the partition between the superconductors (along the
$z$-axis in Fig.1). To illustrate the results obtained by
computing the formula (\ref{charge-term}), we plot the
current-phase diagrams for two different geometries. These
geometries correspond to the different orientations of the
crystals in the right and left sides of the interface
(Fig.1):\newline (i) The basal $ab$-plane in the right side has
been rotated around the $c$-axis by $\alpha $;
$\hat{\mathbf{c}}_{1}\Vert \hat{\mathbf{c}}_{2}$. \newline (ii)
The $c$-axis in the right side has been rotated around the
$b$-axis by $\alpha $; $\hat{\mathbf{b}}_{1}\Vert
\hat{\mathbf{b}}_{2}$.\newline Further calculations require a
certain model of the gap vector (order parameter) $\mathbf{d}$.
\section{Analysis of numerical results}
\label{section4} In the present paper, the nonunitary $f-$wave gap
vector in the $B-$phase (low temperature $T$ and low field $H$) of
superconductivity in $UPt_{3}$ compound has been considered. This
nonunitary bipolar state which explains the weak spin-orbit
coupling in $UPt_{3}$ is \cite{Machida}: \begin{equation}
\mathbf{d}(T,\mathbf{v}_{F})=\Delta
_{0}(T)k_{z}(\hat{\mathbf{x}}\left( k_{x}^{2}-k_{y}^{2}\right)
+\hat{\mathbf{y}}2ik_{x}k_{y}). \label{Model} \end{equation} The
coordinate axes
$\hat{\mathbf{x}},\hat{\mathbf{y}},\hat{\mathbf{z}}$ are selected
along the crystallographic axes
$\hat{\mathbf{a}},\hat{\mathbf{b}},\hat{\mathbf{c}}$ in the left
side of Fig.\ref{fig1}. The function $\Delta _{0}=$$\Delta
_{0}\left( T\right) $ describes the dependence of the gap vector
on the temperature $T$ (our numerical calculations are done at the
low value of temperature $T/T_{c}=0.1$). Using this model of the
order parameter (\ref{Model}) and solution to the Eilenberger
equations (\ref {charge-term}), we have calculated the current
density at the interface numerically. These numerical results are
listed below:\newline 1) The nonunitary property of Green's matrix
diagonal term consists of two parts. The explicit part which is in
the $B$ mathematical expression in Eq.(\ref{charge-term}) and the
implicit part in the $\Omega _{1,2}$ and $\mathbf{d}_{1,2}$ terms.
These $\Omega _{1,2}$ and $\mathbf{d}_{1,2}$ terms are different
from their unitary counterparts. In the mathematical expression
for $\Omega _{1,2}$ the nonunitary mathematical terms
$\mathbf{A}_{1,2}$ are presented. The explicit part will be
present only in the presence of misorientation between gap
vectors,\newline $B=i\mathbf{d}_{1}\times \mathbf{d}_{2}\cdot
(\mathbf{A}_{1}\mathbf{+A}_{2})(\eta \Omega _{2}-\varepsilon
)(\eta \Omega _{1}+\varepsilon )$, but the implicit part will be
present always. So, in the absence of misorientation
$(\mathbf{d}_{1}\mathbf{\Vert d}_{2})$, although the implicit part
of nonunitary exists the explicit part is absent.
\begin{figure}[tbp]
\includegraphics[width=0.9\columnwidth]{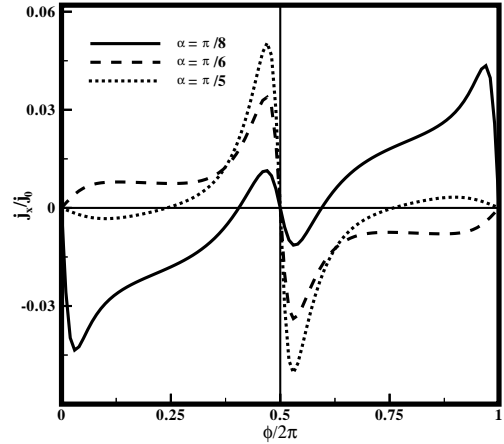}
\caption{Current tangential to the interface versus the phase
difference $\protect\phi $ for the junction between nonunitary
bipolar $f-$wave superconducting bulks, $T/T_{c}=0.15$, geometry
(ii) and the different misorientations ($x$component) .}
\label{fig5}
\end{figure}
This means that in the absence of misorientation, current-phase
diagrams for planar unitary and nonunitary bipolar systems are the
same but the maximum values are  slightly different.\newline 2) A
component of current parallel to the interface $j_{z}$ for
geometry (i) is zero similar to the unitary case \cite{Mahmoodi}
while the other parallel component $j_{x}$ has a finite value (see
Fig.\ref{fig4}). This latter case is a difference between unitary
and nonunitary cases. Because in the junction between unitary
$f-$wave superconducting bulks all parallel components of the
current ($j_{x}$ and $j_{z}$) for geometry (i) are absent
\cite{Mahmoodi}.\newline 3) In Figs.\ref{fig2},\ref{fig3}, the
Josephson current $j_{y}$ is plotted for a certain nonunitary
model of $f-$wave and different geometries. Figs.\ref
{fig2},\ref{fig3} are plotted for the geometry (i) and geometry
(ii) respectively. They are completely unusual and totally
different from their unitary counterparts which have been obtained
in \cite{Mahmoodi}.\newline 4) In Fig.\ref{fig2} for geometry (i),
it is observed that by increasing the misorientation, some small
oscillations appear in the current-phase diagrams as a result of
the non-unitary property of the order parameter, . Also, the
Josphson current at the zero external phase difference $\phi =0$
is not zero but has a finite value. The Josephson current will be
zero at the some finite values of the phase difference.\newline 5)
In Fig.\ref{fig3} for geometry (ii), it is observed that by
increasing the misorientation the new zeros in current-phase
diagrams appear and the maximum value of the current will be
changed non-monotonically.
\begin{figure}[tbp]
\includegraphics[width=0.9\columnwidth]{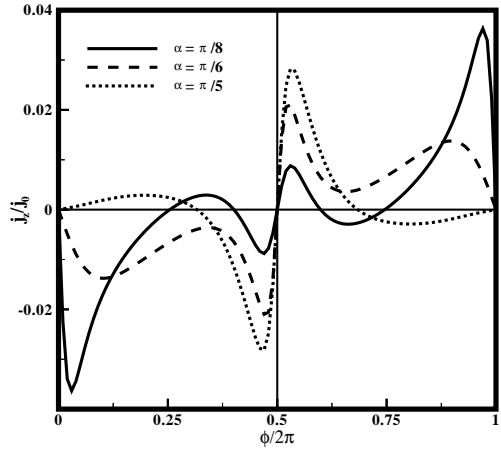}
\caption{Current tangential to the interface versus the phase
difference $\protect\phi $ for the junction between nonunitary
bipolar $f-$wave superconducting bulks, $T/T_{c}=0.15$, geometry
(ii) and the different misorientations ($z-$component).}
\label{fig6}
\end{figure}
In spite of the Fig\ref{fig2} for geometry (i), the Josephson
currents at the phase differences $\phi =0$, $\phi =\pi ,$ and
$\phi =2\pi $ are zero exactly.\newline 6) The current-phase
diagram for geometry (i) and $x-$component (Fig.\ref {fig4}) is
totally unusual. By increasing the misorientation, the maximum
value of the current increases. The components of current parallel
to the interface for geometry (ii) are plotted in Fig.\ref{fig5}
and Fig\ref{fig6}. All the terms at the phase differences $\phi
=0$, $\phi =\pi,$ and $\phi =2\pi $ are zero. The maximum value of
the current-pase diagrams is not a monotonic function of the
misorientation.\newline
\section{Conclusions}
\label{section5} Thus, we have theoretically studied the
supercurrents in the ballistic Josephson junction in the model of
an ideal transparent interface between two misoriented $UPt_{3}$
crystals with nonunitary bipolar $f-$wave superconducting bulks
which are subject to a phase difference $\phi $. Our analysis has
shown that misorientation between the gap vectors creates a
current parallel to the interface and different misorientations
between gap vectors influence the spontaneous parallel and normal
Josephson currents. These have been shown for the currents in the
point contact between two bulks of unitary axial and planar
$f$-wave superconductor in \cite{Mahmoodi} separately. Also, it is
shown that the misorientation of the superconductors leads to a
spontaneous phase difference that corresponds to the zero
Josephson current and to the minimum of the weak link energy in
the presence of the finite spontaneous current. This phase
difference depends on the misorientation angle. The tangential
spontaneous current is not generally equal to zero in the absence
of the Josephson current. The difference between unitary planar
and nonunitary bipolar states can be used to distinguish between
them. This experiment can be used to test the pairing symmetry and
recognize the different phases of $UPt_{3}$.

\end{document}